\documentclass[twocolumn,pre,superscriptaddress]{revtex4-1}

\usepackage{graphicx}
\usepackage{amsmath,amssymb,amsfonts,amsthm}
\usepackage{color}
\usepackage{calrsfs}
\usepackage{epstopdf}

\begin{document}
\title{Sequential Defense Against Random and Intentional Attacks in Complex Networks}

\author{Pin-Yu~Chen}
\affiliation{Department of Electrical Engineering and Computer Science, University of Michigan, Ann Arbor, MI 48109 USA}

\author{Shin-Ming~Cheng}
\affiliation{Department of Computer Science and Information Engineering, National Taiwan University of Science and Technology, Taipei 10607 Taiwan}

\begin{abstract}
Network robustness against attacks is one of the most fundamental researches in network science as it is closely associated with the reliability and functionality of various networking paradigms. However, despite the study on intrinsic topological vulnerabilities to node removals, little is known on the network robustness when network defense mechanisms are implemented, especially for networked engineering systems equipped with detection capabilities. In this paper, a sequential defense mechanism is firstly proposed in complex networks for attack inference and vulnerability assessment, where
the data fusion center sequentially infers the presence of an attack based on the binary attack status reported from the nodes in the network.
The network robustness is evaluated in terms of the ability to identify the attack prior to network disruption under two major attack schemes, i.e., random and intentional attacks.
We provide a parametric plug-in model for performance evaluation on the proposed mechanism and validate its effectiveness and reliability via canonical complex network models and real-world large-scale network topology.
The results show that the sequential defense mechanism greatly improves the network robustness and mitigates the possibility of network disruption by acquiring limited attack status information from a small subset of nodes in the network.
\end{abstract}

\pagestyle{empty}
\maketitle


\section{Introduction}
\label{sec_Intro}
In the past decade, with the advance of computation technology and the accessibility of real-world large-scale network data, the exploration and analysis of  large-scale network attributes have received tremendous attention in network science \cite{Lewis08} as they disclosed the mysterious masks in nature as well as man-made engineered systems and contrive to answer the fundamental networking problems such as network formulation, dependency, resilience and evolution.
Such networks, consisting of numerous nodes and intricate interconnections embedded with heterogeneous network structures in the graph-theoretic point of view, are renowned as \textit{complex networks} \cite{Albert02,Newman03,Wang03}.
Owing to large-scale network size, extreme volume of empirical network data, and potentially biased network sampling techniques \cite{Kurant11}, explicit analysis on the network structure turns out to be computationally infeasible and theoretically intractable. Consequently, collective network attributes instead of exact network topology are preferable for complex network analysis, and the developed measurement metrics (e.g., clustering coefficient and network centrality) play an essential role in network science and they have been applied to aid the design of communication systems \cite{Cui10}. Among all the network attributes, the degree distribution of the entire network is one of the most salient feature that specifies the link characteristics since the degree distribution is defined as the probability distribution of the number of links of an arbitrarily selected node in the complex network, and it can be specified by a few network parameters.

What is of our particular interest in network science is the study of network resilience \cite{Albert00} (i.e., the extent of network tolerance to node removals) because of its kin relation and assessment to network robustness and connectivity in many networked engineering systems \cite{Menth09,Smith11,CPY12,CPY13GlobalSIP,CPY14ComMag}. Typical examples include but are not limited to denial-of-service (DoS) attacks and jamming attacks. In particular, the U.S. Department of Energy (DOE) has identified attack resistance to be one of the seven major properties required for the operation of smart grid \cite{DOE2007}.
From the bird's-eye view of the entire network, the giant connected component vanishes and the entire network is disintegrated into several small components when the fraction of the removed nodes exceeds certain critical value, which is known as the critical phenomenon of percolation theory in statistic physics \cite{Callaway00}. More importantly, this critical phenomenon can be well mapped to the network robustness and connectivity of many practical networked engineering systems, owing to the network resilience protocols that the network retains its operations as long as a majority of nodes remain its functionality (i.e., most of the nodes are still connected). Throughout this paper, the critical phenomenon for network disruption caused by node removals are used to evaluate the performance of the proposed network defense mechanism and we denote the critical value for network disruption as the \textit{percolation-based connectivity}.

\begin{figure}[t]
    \centering
    \includegraphics[width=3in]{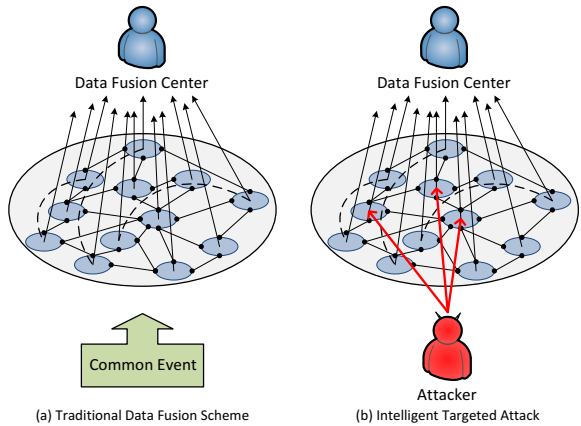}
    \caption{Structure of practical engineering system. A data fusion center is responsible for data inference and decision making based on the feedback data from the network. The solid lines represent localized connections (e.g., physical links in a power grid) and the dashed lines represent delocalized connections (e.g., friends in a social network) in the complex network.
    (a) Traditional data fusion scheme. Each node feedbacks its observation on a common event (e.g., channel vacancy or temperature) to the data fusion center for hypothesis test. (b) Intelligent targeted attack. Red solid arrows point to the targeted nodes.
    An intelligent adversary leverages the network topology to target the most vulnerable nodes to disrupt the entire network. As this targeted attack is not a common event to all the nodes, most of the nodes are unaware of the attack and therefore it is more difficult to be detected. Consequently, intelligent targeted attack hinders the attack inference precision and poses severe threats on the network robustness.}
    \label{fig_system_1}
\end{figure}

Our physical model is built upon the structure of many practical networked engineering systems where a data fusion is responsible for data inference and decision making as illustrated in Fig. \ref{fig_system_1}.
Although a vast amount of research has been done in analyzing intrinsic network resilience in complex networks and devising efficient intrusion/anomaly detection techniques in practical networked engineering systems separately, a complete and interdisciplinary network robustness analysis including both the intrinsic network resilience as well as the embedded attack detection capability is still poorly understood.  In this paper, a sequential defense mechanism is first proposed in complex networks where each node performs individual attack detection and sequentially reports binary attack status (i.e., under attack or not) to the data fusion center as shown in Fig. \ref{fig_system_1}. The data fusion center then sequentially infers the presence of network attacks based on the feedback and makes a final decision when sufficient information has been collected. This mechanism is particularly applicable to networking paradigms with enormous number of nodes and stringent data transmission resources. It is also worth mentioning that the proposed sequential defense mechanism is quite distinct from the traditional data fusion scheme \cite{Varshney96} due to the fact that the network attack may not be a common event to all the nodes in the network as illustrated in Fig. \ref{fig_system_1}. In other words, an intelligent adversary can target at some crucial nodes instead of launching attacks on the entire network to efficiently disrupt the network and reduce the risks of being detected, which therefore hinders the attack inference precision and poses severe threats on the network robustness.


The performance of the proposed sequential defense mechanism is evaluated under random and intentional attacks, as random attack plays an identical role of temporal node disfunction and intentional attack refers to malicious attack caused by an adversary.
We provide a parametric plug-in model for performance evaluation on the sequential defense mechanism, and we implement our mechanism in both canonical complex network models and empirical network data to validate its reliability and effectiveness. In addition to analyzing the critical value to sustain percolation-based connectivity via statistic physics approaches \cite{Callaway00}, we would like to point out that our defense mechanism is a general framework which does not depend on any underlaying complex network models but is applicable to any network with arbitrary network structures, provided that the critical value of the network can be realized at hand. The results show that our defense mechanism greatly enhances the network robustness and provides reliable protection against fatal attacks, even in the complex networks with fragile network structure and weak detection capability, which also offer new insights toward network robustness enhancement and robust network design.

The rest of this paper is organized as follows. The related works are summarized in Sec. \ref{sec_Related}. Preliminaries on the percolation-based connectivity and the canonical complex network models are introduced in Sec. \ref{sec_Preliminary}. The system model and sequential defense mechanism are elucidated in Sec. \ref{sec_system}. The critical values under random and intentional attacks are analyzed in Sec. \ref{sec_attack}. The analysis on the sequential defense mechanism is derived in Sec. \ref{sec_performance_analysis}. The performance evaluation of the proposed defense mechanism in canonical complex network models and empirical network data are shown in Sec. \ref{sec_performance}. Sec. \ref{sec_robust} provides discussions for robust network design. Finally, Sec. \ref{sec_con} concludes this paper.

\section{Related Works}
\label{sec_Related}
The intrinsic topological vulnerabilities regarding different network structures under random and intentional attacks were first introduced in \cite{Albert00}. Compared with random attack, intentional attack is shown to be quite effective in disintegrating the entire network by removing a relatively small fraction of nodes with the highest degree in the network. As many real-world networks are observed to possess a heavy-tailed degree distribution, such as the webpage links in World Wide Web (WWW) \cite{Barabasi99}, router maps in Internet \cite{Faloutsos99} and contacts in email networks \cite{Ebel02}, the existence of nodes with a relatively large number of links render such networks particularly vulnerable to intentional attack. Moreover,
it has been demonstrated in \cite{Xiao08} that intentional attack is the most effective attack strategy to disrupt the entire network when the network topology is known by the adversary, which suggests intentional attack to be an ever-increasing threat on the network robustness of many networked engineering systems.

With the aid of statistic physics and percolation theory \cite{Callaway00}, the critical values (i.e., the fraction of removed nodes) for a complex network to sustain random and intentional attacks prior to network disruption are investigated in \cite{Cohen00} and \cite{Cohen01}, respectively, which offer analytically tractable tools for network robustness assessment. Please note that most of the existing research on network robustness against attacks mainly focus on intrinsic topological vulnerabilities while the impacts of implementing network defense mechanisms on the network robustness are still poorly understood.
A naive perfect node protection scheme is proposed in \cite{Xiao11} to prevent a subset of nodes in the network from being attacked, which can be shown as a degenerate case of our proposed model.
A two-player, zero-sum attack and defense game is introduce in \cite{CPY11,CPY12,CPY14JIOT} to alleviate the damage caused by intentional attack by acquiring attack status from each node for attack inference and defense reaction, and the outcome of the game equilibrium is used to evaluate the network robustness. However, this mechanism is not suitable in networked systems with an enormous number of nodes and stringent data transmission resources as frequent data transmissions may deteriorate the system performance and inevitably incur excessive energy consumption.

To provide efficient defense for complex networks, a sequential hypothesis test approach \cite{Wald} is proposed to identify the attack while acquiring as little information from the network as possible. The data fusion center acquires the reports from each node in descending degree order, and therefore it is able to spare the transmissions of the unreported nodes once the process of sequential test terminates, which balances the goals of promptness and accuracy for attack inference.

\section{Preliminaries on Complex Networks}
\label{sec_Preliminary}
\subsection{Percolation-based Connectivity in Complex Networks}
In the realms of network science, the degree (the number of links of a node) distribution plays an essential role in characterizing the collective topological features. With the advance of computation capability and the accessibility of large-scale network data, the long-believed totally random link connections \cite{ER59} have been overthrown by the extraordinary and ubiquitous degree distributions found in a variety of research areas, such as the power-law distribution in the Internet router-level topological maps \cite{Faloutsos99} and the small world phenomenon in social networks \cite{Watts98}. We denote the degree distribution of a complex network by $P(k)$, where $k \in [k_{min},k_{max}]$ and $k_{min}$ ($k_{max}$) is the smallest (largest) degree of the complex network.
From the bird's eye view, the network attack can be mapped to the node removal in the corresponding network graph (all links attached to the removed node are removed as well), and the network is said to be connected in percolation sense if the giant component (the connected component that includes a majority of nodes) still exists after node removal, which we refer to as the percolation-based connectivity. The physical interpretation of the percolation-based connectivity is that owing to the network resilience protocols \cite{Smith11}, the network can continue its main operations under temporal node disfunction as long as most of the nodes are still connected.

According to the seminal work in~\cite{Molloy95}, given the degree distribution $P(k)$ of an arbitrary network, a giant component containing the majority of the nodes exists in the network if $P(k)$ satisfies the criterion $\sum_k k(k-2)P(k)>0$, which is equivalent to the condition
\begin{align}
\label{eqn_condition}
\tau \triangleq \frac{\mathbb{E}[\mathbf{K}^2]}{\mathbb{E}[\mathbf{K}]}>2,
\end{align}
where $\mathbf{K} \in [k_{min}, k_{max}]$ is the random variable representing the degree of a randomly selected node.
With the aid of percolation theory, the critical phenomenon of network disruption occurs if more than $q_c$ fraction of nodes are removed from the network, where the critical value $q_c$ can be estimated when the remaining degree distribution satisfies the criterion $\tau_c=2$. In other words, the complex network transitions from the connected phase to the disconnected phase in percolation sense once more than $q_c$ fraction of nodes are removed. Throughout this paper, the critical value $q_c$ is used to evaluate the network robustness under different network structures and attack schemes. Please note that in the case of small-scale networks, the critical value can be obtained by performing exhaustive node removal experiments (i.e., searching over all possible node removal strategies) instead of using statistic physics approaches (i.e., estimating $q_c$ by degree distribution).

\subsection{Canonical Complex Network Models}
In this paragraph, we introduce three canonical complex network models that serve as the platforms for performance evaluation of the proposed defense mechanism.
\begin{itemize}
  \item \textbf{ER network}. In an ER network \cite{ER59}, a link between any arbitrarily selected node pair is present with probability $p_{ER}$. If the network size is large enough, the degree distribution approaches to the Poisson distribution
      $P(k)=e^{-\widehat{k}} \frac{\widehat{k}^k}{k!}$, where $\widehat{k}=N \cdot p_{ER}$ is the mean degree of the network and $N$ is the number of nodes in the network.
  \item \textbf{Power-law network}. A power-law network possesses a skewed degree distribution $P(k) \sim k^{-\alpha}$, where $\alpha>0$ is the skewness parameter.  The heavy tail of the degree distribution suggests the existence of the hub nodes that few nodes have relatively high degree compared with most of the nodes in the network, which well explains the connectivity of the WWW \cite{Barabasi99} or the Internet router maps \cite{Faloutsos99}.
  \item \textbf{Exponential network}. An exponential network has its degree distribution $P(k) \sim \frac{1}{\beta}e^{-\frac{k}{\beta}}$, where $\beta>0$ is the mean degree of the network in the large scale network limit.
      It is demonstrated in \cite{Sole08} that the degree distribution of the power grid can be characterized by the exponential distribution, both in the national power grid scale and the European power grid scale.
 \end{itemize}

\section{System Model}
\label{sec_system}
\subsection{Network Model and Sequential Defense Mechanism}
\label{subsec_system_network_model}
Without loss of generality, we consider the complex network model consisting of $N$ nodes characterized by its degree distribution $P(k)$ and the corresponding critical value $q_c$ to sustain network connectivity against attacks. Each node is equipped with certain detection capability, for instance, intrusion detection techniques \cite{Mukherjee94} or malicious activity filtering \cite{Androulidaki09} for attack inference. The $N$ nodes are sorted in descending degree order, i.e., $k_1 \geq k_2, \geq \ldots \geq k_N$. Let $H_1$ denote the hypothesis that the attack occurs in the complex network (either on one node or several nodes simultaneously), and $H_0$ denote the alternative hypothesis of a null attack (i.e., there is no attack in the network). Based on the nodal detection, every node sequentially reports its binary hypothesis testing decision to the data fusion center in descending degree order since intuitively the removal of nodes with higher degree results in more severe damage to the network robustness.

It is worth mentioning that although enabling local information exchange or cooperative detection among nodes may enhance the attack inference precision, these approaches inevitably increase the computation and data transmission overheads. Throughout this paper, we will concentrate on the degree-based sequential defense mechanism owing to its feasibility and simplicity.
In practice, these local decisions can be transmitted in the header of data packets, or a node is regarded as being attacked if it fails to reply to the periodic beaconing from the data fusion center.
If the attack is confirmed by the data fusion center, network defense schemes such as node quarantine or system renewal will be launched to alleviate the damage, otherwise it keeps surveillance on the collected information.

The advantages of reporting binary attack status for sequential defense in complex networks reside in the feasibility of data transmission and computation complexity in the large-scale networked systems. The enormous network size (e.g., Internet routers or wireless sensors) render simultaneous data transmissions infeasible, especially for wireless networks with scarce radio resources. Moreover, owing to the large network size and limited computational power, analyzing the collected information from all nodes incurs tremendous computation overheads and it may fail to provide timely defense. Consequently, sequential hypothesis test with minimum (one-bit) feedback information is an essential must for attack inference in complex networks because of its least additional communication overheads and timely defense. In other words, the sequential hypothesis test terminates once sufficient information is collected and a final decision is made by the data fusion center so that the system can spare the transmissions of the unreported nodes.

Let $x_i$ denote the attack status reported by the $i$th node. $x_i=1$ when the attack on the $i$th node is detected and $x_i=0$ for null attack on the $i$th node. We assume that the detection capability of each node is identical with probability of detecting an attack $P_D$ and probability of false alarm $P_F$. Each node performs independent hypothesis test such that the joint probability distribution of the first $m \leq N$ reports when $H_j$ is true can be represented as $P(x_1,\ldots,x_m|H_j)=\prod_{i=1}^{m} P(x_i|H_j)$.
In general, we assume $P_D \geq P_F$, otherwise the sequential hypothesis test should be altered for attack inference.

\subsection{Sequential Probability Ratio Test (SPRT)}
\label{subsec_system_SPRT}
Using sequential analysis \cite{Wald}, let $P_{jm}=\prod_{i=1}^{m} P(x_i|H_j)$ denote the probability of obtaining a report sample ($x_1,\ldots,x_m$), the sequential hypothesis test is carried out by performing the probability ratio test with two specified parameters $A$ and $B$. After receiving the report from the $m$th node, if $\frac{P_{1m}}{P_{0m}} \geq A~\left(\frac{P_{1m}}{P_{0m}} \leq B \right)$, then the data fusion center declares the presence of a (null) attack, otherwise it keeps surveillance on the next report for attack inference. Moreover, for purposes of practical computation, it is much more convenient to perform sequential probability ratio test by computing the logarithm of the ratio $\frac{P_{1m}}{P_{0m}}$ instead of the ratio itself as the product of individual tests can be decomposed into sum of the log likelihood ratios.

Let $z_i=\log \frac{P(x_i|H_1)}{P(x_i|H_0)}$ denote the log likelihood ratio of the $i$th report and $\Lambda_m=\sum_{i=1}^{m} z_i= \sum_{i=1}^{m} \log \frac{P(x_i|H_1)}{P(x_i|H_0)}$ denote the cumulative value of the first $m$ reports for hypothesis testing. Consequently, the sequential hypothesis test is terminated with hypothesis $H_1$ ($H_0$) if $\Lambda_m \geq \log A$ ($\Lambda_m \leq \log B$), otherwise the process is continued by taking an additional report. These two parameters $A$ and $B$ can be determined by setting $A=\frac{1-\theta}{\delta}$ and $B=\frac{\theta}{1-\delta}$, where $\delta=P(\textnormal{say}~H_1~\textnormal{when}~H_0~\textnormal{is~true})$ and $\theta=P(\textnormal{say}~H_0~\textnormal{when}~H_1~\textnormal{is~true})$ are the required false alarm and miss detection probabilities at the system level.

\section{Critical Values under Random and Intentional Attacks}
\label{sec_attack}
Incorporating the topological vulnerabilities of the complex network, the critical value $q_c$ to sustain percolation-based connectivity under random and intentional attacks are analyzed with respect to distinct canonical complex network models. For comprehensive analysis and clear reading, only the results are displayed in this section and the mathematical derivations are placed in the appendices.
\subsection{Random Attack}
Random attack on the $q$ fraction of nodes in the network plays an identical role of random node removal. Given the original network degree $\mathbf{K_0}$ of a randomly selected node, the critical value for random attack becomes
\begin{align}
\label{eqn_random_attack_thre}
q_c^{ran}=1-\frac{1}{\tau_0 -1},
\end{align}
where $\tau_0 \triangleq \frac{\mathbb{E}[\mathbf{K_0}^2]}{\mathbb{E}[\mathbf{K_0}]}$ is calculated from the original degree distribution.
For ER network, $q_{c}^{ran-ER}=1-\frac{1}{\widehat{k}}$.
For power-law network, $q_{c}^{ran-POW}=1-\left(\frac{2-\alpha}{3-\alpha}\frac{k_{1}^{3-\alpha}- k_{N}^{3-\alpha}}{k_{1}^{2-\alpha}- k_{N}^{2-\alpha}}-1\right)^{-1}$.
For exponential network, $q_c^{ran-EXP}=1-\left({\frac{k_N^2+2 k_N \beta+2 \beta^2}{k_N+\beta}-1}\right)^{-1}$.
Detailed derivations can be found in Appendix \ref{appex_critical_random}.

\subsection{Intentional Attack}
As demonstrated in \cite{Cohen01}, removing $q$ fraction of nodes with the highest degree in the network is equivalent to randomly removing $\widetilde{q}$ fraction of nodes in the remaining network with new cutoff degree $\widetilde{k}_{max}<k_{max}$. With the continuous degree approximation and the relation $\sum_{k_{max}}^{\infty} P(k)=\int_{k_{max}}^{\infty} P(k)dk=\frac{1}{N}$,
the new cutoff degree $\widetilde{k}_{max}$ can be evaluated from
\begin{align}
\label{eqn_intentional_removal}
\int_{k=\widetilde{k}_{max}}^{k_{max}} P(k) dk=\int_{k=\widetilde{k}_{max}}^{\infty} P(k)dk -\frac{1}{N}=q.
\end{align}
Moreover, $\widetilde{q}$ can be interpreted as the link deletion probability of a randomly selected link leading to a deleted node, which equals the ratio of the number of links belonging to the deleted nodes to the number of links \cite{Cohen01,Newman01}, i.e., $\widetilde{q}=\sum_{k=\widetilde{k}_{max}}^{k_{max}} \frac{k P(k)}{\mathbb{E}[\mathbf{K_0}]}$. By specifying the relations between link deletion probability and targeted node removal in (\ref{eqn_intentional_removal}), the critical value $q_c$ under intentional attack can be obtained by evaluating the critical link deletion probability $\widetilde{q}_c$ with the ubiquitous criterion for percolation-based connectivity in (\ref{eqn_random_attack_thre}).

For ER network, $q_c^{int-ER}=\frac{1}{N}- e^{-\widehat{k}} \frac{\widehat{k}^{\widetilde{k}_{max}-1}}{\left( \widetilde{k}_{max}-1 \right)!}+1-\frac{1}{\widehat{k}}$.
For power-law network, $q_c^{int-POW}=\left(\frac{\widetilde{k}_{max}}{k_N}\right)^{1-\alpha}$.
For exponential network, the critical value can be obtained by solving
\begin{align}
\label{eqn_intentional_EXP_main}
&\left[ 1-\ln \left( q_c^{int-EXP}+ \frac{1}{N}\right) \right]\left(q_c^{int-EXP}+\frac{1}{N}\right) \nonumber \\
&+\frac{k_N+\beta}{k_N^2+2 k_N \beta+2 \beta^2-k_N-\beta}-1=0.
\end{align}
Detailed derivations can be found in Appendix \ref{appex_critical_intentional}.

\subsection{Unified Notations for Attack Schemes and SPRT}
Since each node sequentially reports its one-bit detection result to the data fusion center for attack inference,
let $a_i$ denote the probability of attacking $i$th node, we introduce the unified notations for the aforementioned attack schemes as
\begin{align}
\label{eqn_attack_coefficient_formulation}
P(x_i|H_j)=
\left\{
  \begin{array}{ll}
      \mathcal{B}(a_i \cdot P_D),~&\textnormal{if}~j=1, \\
      \mathcal{B}(P_F),~&\textnormal{if}~j=0,
  \end{array}
\right.
\end{align}
where $\mathcal{B}(p)$ is the Bernoulli trial with probability of success ($x_i=1$) equals $p$.
Incorporating the attack schemes, we have
\begin{align}
\label{eqn_attack_coefficient}
a_i^{ran}&=q^{ran},~\forall~i;~  \\
a_i^{int}&=\mathbf{1}_{i \leq \lceil Nq^{int} \rceil}+\frac{P_F}{P_D}\mathbf{1}_{i > \lceil Nq^{int} \rceil} \nonumber \\
&=(1-\frac{P_F}{P_D}) \mathbf{1}_{i \leq \lceil Nq^{int} \rceil}+\frac{P_F}{P_D},
\end{align}
where $\mathbf{1}_E$ denotes the indicator function of the event $E$ and $\lceil x \rceil$ is the smallest integer that exceeds $x$.

\section{Performance Analysis of Sequential Defense in Complex Networks}
\label{sec_performance_analysis}
Given the specified system parameters ($\delta,\theta$), we are interested in the effectiveness and the performance of the proposed sequential defense mechanism against random and intentional attacks in complex networks. Furthermore, knowing the critical value $q_c$, the data fusion center is required to infer the presence of the attack prior to the network disruption. In other words, a final decision has to be made according to the first $M_c=\lceil N \cdot q_c \rceil$ reports for practical implementation purposes, which we refer to as the worst case scenario. Upon the reception of the $M_c$th report, if a final decision has not been reached, the data fusion center declares the presence of attack when $0<\Lambda_{M_c}<\log A$ and declares a null attack when $\log B<\Lambda_{M_c} \leq 0$.

Let $M_j$ denote the expected number of reports required for hypothesis testing when $H_j$ is true.
The proposed sequential defense mechanism is regarded as effective against attacks in the complex network if $M_1 \leq M_c$, i.e., the number of reports required for attack inference is less than the threshold of network disruption, otherwise the defense is in vain since it fails to provide timely defense reaction. We derive the closed-form expressions of $M_1$ for random and intentional attacks, and we prove that for intentional attack, taking additional reports from $m > M_c$ nodes does not improve the performance of the sequential defense mechanism.

\subsection{Random Attack}
\label{subsec_performance_random}
For random attack, with (\ref{eqn_attack_coefficient_formulation}) we have
\begin{align}
P_{1m}&=\prod_{i=1}^m P(x_i|H_1) \nonumber \\
&=(q^{ran}P_D)^{d_m}(1-q^{ran}P_D)^{m-d_m}; \\
P_{0m}&=\prod_{i=1}^m P(x_i|H_0) \nonumber \\
&=(P_F)^{d_m}(1-P_F)^{m-d_m},
\end{align}
where $d_m$ is the number of ones in the first $m$ reports. Simple calculation on $\Lambda_m$ yields
\begin{align}
\label{eqn_performance_random}
\Lambda_m=d_m \log \frac{q^{ran}P_D}{P_F}+(m-d_m) \log \frac{1-q^{ran}P_D}{1-P_F}.
\end{align}
Following the process of SPRT in Sec. \ref{subsec_system_SPRT}, the sequential defense criterion for random attack becomes
\begin{align}
\left\{
  \begin{array}{ll}
      \textnormal{say}~H_1,~\textnormal{if}~d_m  \geq  \frac{\log A}{\log\frac{q^{ran}P_D}{P_F}-\log\frac{1-q^{ran}P_D}{1-P_F}} \nonumber \\
      ~~~~~~~~~~~~~~~~~~+m \frac{\log\frac{1-P_F}{1-q^{ran}P_D}}{\log \frac{q^{ran}P_D}{P_F}-\log \frac{1-q^{ran}P_D}{1-P_F}}, \\
      \textnormal{say}~H_0,~\textnormal{if}~d_m  \leq  \frac{\log B}{\log\frac{q^{ran}P_D}{P_F}-\log\frac{1-q^{ran}P_D}{1-P_F}} \nonumber \\
      ~~~~~~~~~~~~~~~~~~+m \frac{\log\frac{1-P_F}{1-q^{ran}P_D}}{\log \frac{q^{ran}P_D}{P_F}-\log \frac{1-q^{ran}P_D}{1-P_F}}, \\
      \textnormal{keep surveillance},~\textnormal{otherwise}.
  \end{array}
\right.
\end{align}
The expected number of reports to identify random attack when $H_1$ is true is
\begin{align}
\label{eqn_performance_random_M1}
M_1^{ran}&=\frac{\theta \log B+(1-\theta) \log A}{\mathbb{E}[z_i|H_1]}  \nonumber \\
&=\frac{\theta \log \frac{\theta}{1-\delta}+(1-\theta) \log \frac{1-\theta}{\delta}}{q^{ran}P_D\log \frac{q^{ran}P_D}{P_F} + (1-q^{ran}P_D)\log \frac{1-q^{ran}P_D}{1-P_F}}.
\end{align}

For the worst case scenario, if $m$ is large enough, from central limit theorem we obtain the lower bounds of the probability that the SPRT will terminate by declaring attack or null attack with $m \leq M_c$ reports as \cite{Wald}
\begin{align}
\label{eqn_performance_random_worst_1}
&P(\textnormal{declare~attack})=P(\Lambda_m \geq \log A) \geq 1-\Phi(y_1(M_c)); \\
\label{eqn_performance_random_worst_2}
&P(\textnormal{declare~null~attack})=P(\Lambda_m \leq \log B) \geq \Phi(y_2(M_c)),
\end{align}
where $\Phi(x)$ is the cumulative density function (CDF) of a standard normal distribution, and
\begin{align}
y_1(M_c)&=\frac{\log A-M_c \mathbb{E}[z_i|H_1]}{\sqrt{M_c}\sigma(z_i|H_1)}; \\
y_2(M_c)&=\frac{\log B - M_c \mathbb{E}[z_i|H_0]}{\sqrt{M_c} \sigma(z_i|H_0)};\\
\mathbb{E}[z_i|H_0]&=P_F \log \frac{q^{ran}P_D}{P_F}+(1-P_F) \log \frac{1-q^{ran}P_D}{1-P_F}; \\
\sigma(z_i|H_1)&=\sqrt{q^{ran}P_D(1-q^{ran}P_D)} \log \frac{q^{ran}P_D(1-P_F)}{P_F(1-q^{ran}P_D)};\\
\sigma(z_i|H_0)&=\sqrt{P_F(1-P_F)} \log \frac{q^{ran}P_D(1-P_F)}{P_F(1-q^{ran}P_D)},
\end{align}
where
$\sigma(z_i|H_1)$ and $\sigma(z_i|H_0)$ are the standard deviation of $z_i$ under $H_1$ and $H_0$, respectively.
Moreover, when a final decision needs to be made upon the reception of $M_c$th report, the system level false alarm and miss detection probabilities ($\delta,\theta$) when taking $M_c$ reports are bounded by \cite{Wald}
\begin{align}
\label{eqn_performance_random_worst_3}
\delta(M_c)&\leq \delta+\Phi(y_3(M_c))-\Phi(y_4(M_c));  \\
\label{eqn_performance_random_worst_4}
\theta(M_c)&\leq \theta+\Phi(y_5(M_c))-\Phi(y_6(M_c)),
\end{align}
where
\begin{align}
y_3(M_c)&=\frac{\log A - M_c \mathbb{E}[z_i|H_0]}{\sqrt{M_c}\sigma(z_i|H_0)};\\ y_4(M_c)&=-\sqrt{M_c}\frac{\mathbb{E}[z_i|H_0]}{\sigma(z_i|H_0)};\\ y_5(M_c)&=-\sqrt{M_c}\frac{\mathbb{E}[z_i|H_1]}{\sigma(z_i|H_1)};\\
y_6(M_c)&=\frac{\log B - M_c \mathbb{E}[z_i|H_1]}{\sqrt{M_c}\sigma(z_i|H_1)}.
\end{align}

The aforementioned equations are well-known results from \cite{Wald} applied by the specified parameters $q^{ran}$, $P_F$, $P_D$ and $M_c$. Interested readers are referred to \cite{Wald} for more details.

\subsection{Intentional Attack}
\label{subsec_performance_intentional}
Let $d_m$ denote the number of nodes reporting attack for the first $M=\lceil Nq^{int} \rceil$ reports and $d_{m^{\prime}}$ denote the number of nodes reporting attack starting from the $m^{\prime}$th node ($m^{\prime}>M$). With (\ref{eqn_attack_coefficient}), we obtain
\begin{align}
\label{eqn_performance_intentional_product}
P_{1m}&=P_D^{d_m}\left[(1-P_D)^{m-d_m}\mathbf{1}_{m \leq M} + (1-P_D)^{M-d_m}\mathbf{1}_{m > M}\right] \nonumber \\
&~~\cdot \left[ \mathbf{1}_{m \leq M} + P_F^{d_{m^{\prime}}}(1-P_F)^{m-M-d_{m^{\prime}}} \mathbf{1}_{m>M}\right] \nonumber \\
&=P_D^{d_m}\left[(1-P_D)^{m-d_m}\mathbf{1}_{m \leq M} +  (1-P_D)^{M-d_m} \right. \nonumber \\
&~~\cdot \left. P_F^{d_{m^{\prime}}}(1-P_F)^{m-M-d_{m^{\prime}}} \mathbf{1}_{m>M}         \right], \\
P_{0m}&=P_F^{d_m}\left[(1-P_F)^{m-d_m}\mathbf{1}_{m \leq M} +  (1-P_F)^{M-d_m} \right. \nonumber \\
&~~\cdot \left.P_F^{d_{m^{\prime}}}(1-P_F)^{m-M-d_{m^{\prime}}} \mathbf{1}_{m>M} \right].
\end{align}
The cumulative log likelihood ratio becomes
\begin{align}
\label{eqn_performance_intentional_log_likelihood}
\Lambda_m&=d_m \log \frac{P_D}{P_F}+\left[(m-d_m) \log \frac{1-P_D}{1-P_F}\right]\mathbf{1}_{m \leq M} \nonumber \\
&~~+\left[(M-d_m)\log\frac{1-P_D}{1-P_F}+d_{m^{\prime}} \log \frac{P_F}{P_F} \right. \nonumber \\
&~~+ \left.(m-M-d_{m^{\prime}}) \log \frac{1-P_F}{1-P_F}\right] \mathbf{1}_{m>M} \nonumber \\
&=d_m \log \frac{P_D}{P_F}+\left[(m-d_m) \log \frac{1-P_D}{1-P_F}\right]\mathbf{1}_{m \leq M} \nonumber \\
&~~+ (M-d_m)\left[\log\frac{1-P_D}{1-P_F}\right]\mathbf{1}_{m>M} \nonumber \\
&=d_m \left( \log \frac{P_D}{P_F}-\log \frac{1-P_D}{1-P_F}\right)  \nonumber \\
&~~+\left(m \mathbf{1}_{m \leq M}+ M \mathbf{1}_{m > M} \right)\log \frac{1-P_D}{1-P_F},
\end{align}
which suggests that taking additional reports starting from the $m^{\prime}$th node does not help to improve the performance of the sequential defense mechanism as intuitively intentional attack targets only on the first $M$ nodes. The sequential defense criterion for intentional attack is
\begin{figure}[t]
    \centering
    \includegraphics[width=3in]{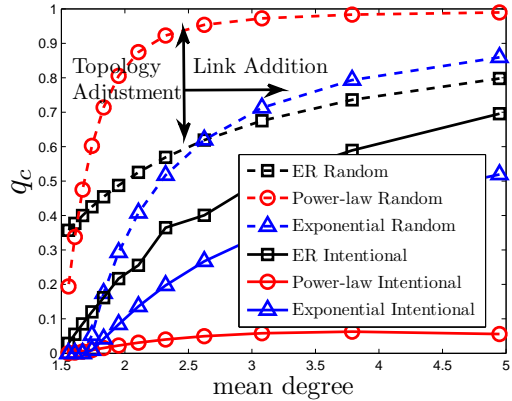}
    \caption{Vulnerabilities of different canonical complex network models under random and intentional attacks. Intentional attack is much more effective in disintegrating a complex network compared with random attack. Although power-law networks are resilient to random attack, they are very vulnerable to intentional attack due to the existence of hub nodes with relatively high degree.}
    \label{fig_qc_mean_deg_2}
\end{figure}

\begin{figure}[t]
    \centering
    \includegraphics[width=3in]{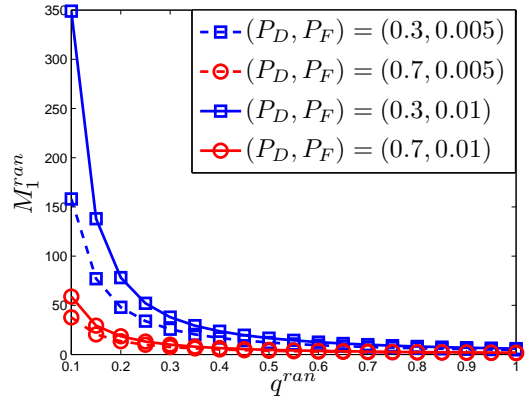}
    \caption{Number of reports required for attack detection ($M_1^{ran}$) with respect to $q^{ran}$ under random attack. $M_1^{ran}$ is shown to be a decreasing function of $P_D$ due to better precision in attack inference. $M_1^{ran}$ increases with $P_F$ to distinguish between attack and null attack.}
    \label{fig_M1_ran_q_3}
\end{figure}
\begin{align}
\left\{
  \begin{array}{ll}
      \textnormal{say}~H_1,~\textnormal{if}~d_m  \geq \frac{\log A}{\log\frac{P_D}{P_F}-\log\frac{1-P_D}{1-P_F}} \nonumber \\
      ~~~~~~~~~~~~~~~~~~+\left(m \mathbf{1}_{m \leq M} + M \mathbf{1}_{m > M} \right) \frac{\log\frac{1-P_F}{1-P_D}}{\log \frac{P_D}{P_F}-\log \frac{1-P_D}{1-P_F}}, \\
      \textnormal{say}~H_0,~\textnormal{if}~d_m  \leq \frac{\log B}{\log\frac{P_D}{P_F}-\log\frac{1-P_D}{1-P_F}} \nonumber \\
      ~~~~~~~~~~~~~~~~~~+\left(m \mathbf{1}_{m \leq M} + M \mathbf{1}_{m > M} \right) \frac{\log\frac{1-P_F}{1-P_D}}{\log \frac{P_D}{P_F}-\log \frac{1-P_D}{1-P_F}}, \\
      \textnormal{keep surveillance},~\textnormal{otherwise}.
  \end{array}
\right.
\end{align}
The expected number of reports required to identify intentional attack when $H_1$ is true is
\begin{align}
\label{eqn_performance_intentional_M1}
M_1^{int}=\frac{\theta \log \frac{\theta}{1-\delta}+(1-\theta) \log \frac{1-\theta}{\delta}}{P_D\log \frac{P_D}{P_F} + (1-P_D)\log \frac{1-P_D}{1-P_F}}.
\end{align}

Consequently, the sequential defense mechanism loses its appeals if $M_1>M_c$ since the reports received are insufficient for attack inference before the adversary disrupts the entire network as proved in (\ref{eqn_performance_intentional_log_likelihood}).
Moreover, it is easy to show that the performance of worst case scenario ($M=M_c$) for intentional attack is identical to that of random attack by substituting $M_c=\lceil Nq_c^{int} \rceil$ and $a_i^{int}$ into (\ref{eqn_performance_random_worst_1}), (\ref{eqn_performance_random_worst_2}), (\ref{eqn_performance_random_worst_3}) and (\ref{eqn_performance_random_worst_4}).

\section{Performance Evaluation}
\label{sec_performance}
\begin{figure}[t]
    \centering
    \includegraphics[width=3in]{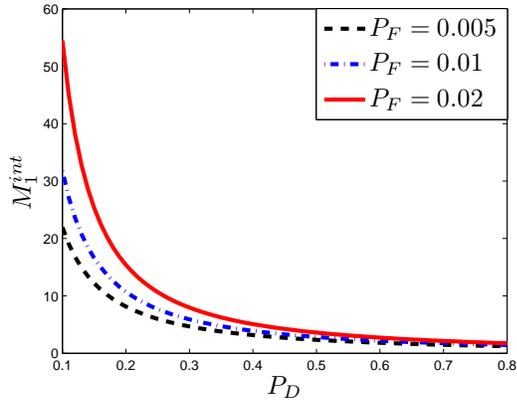}
    \caption{Number of reports required for attack detection ($M_1^{int}$) with respect to $P_D$ under intentional attack. Compared with random attack, the proposed sequential defense mechanism requires only a few number of reports to target intentional attack, even in the low detection probability regime.}
    \label{fig_M1_int_PD_4}
\end{figure}

\begin{figure}[t]
    \centering
    \includegraphics[width=3in]{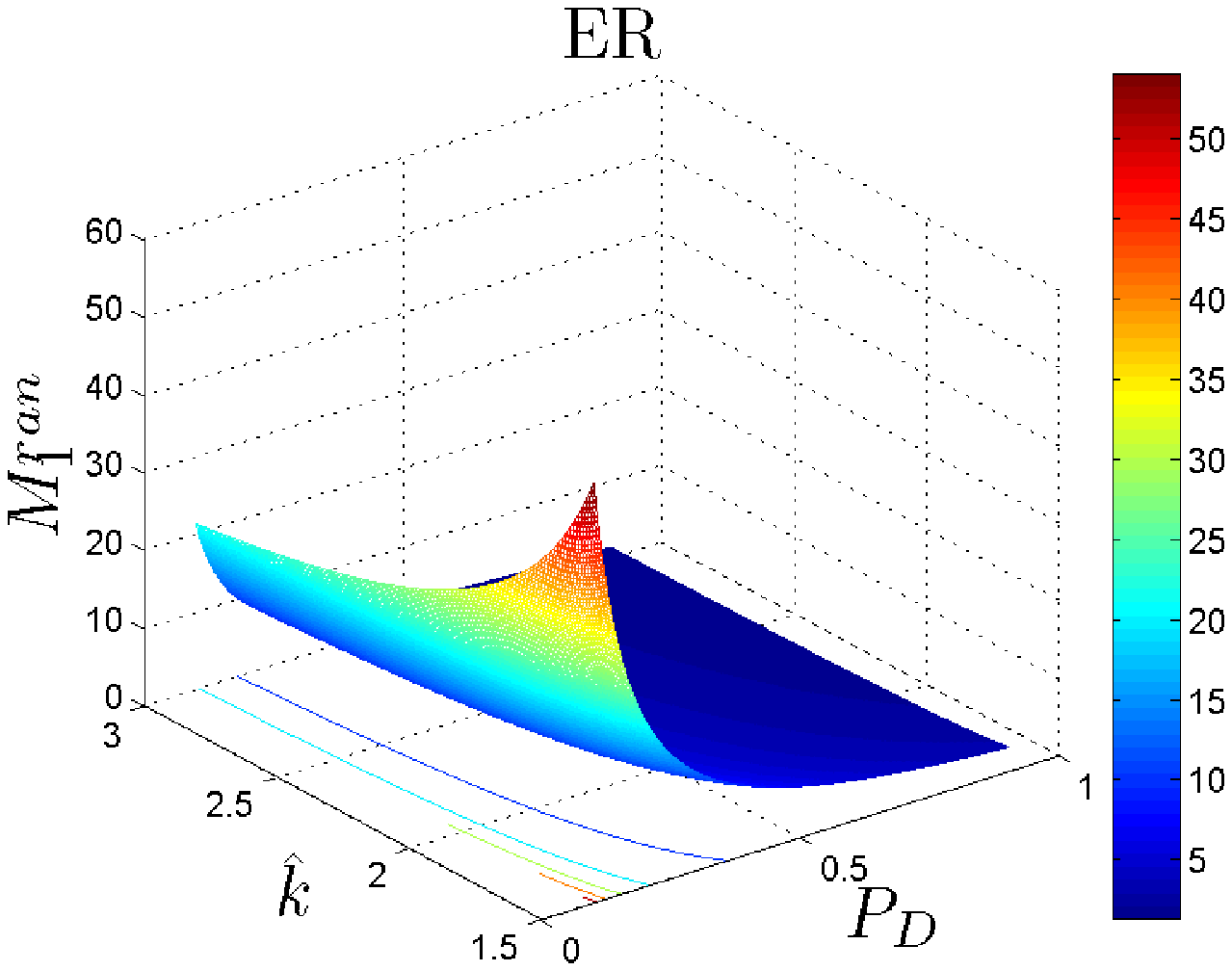}
    \caption{Performance of sequential defense against random attack in ER networks with $P_F=0.001$.}
    \label{fig_ER_3D_5}
\end{figure}

\begin{figure}[t]
    \centering
    \includegraphics[width=3in]{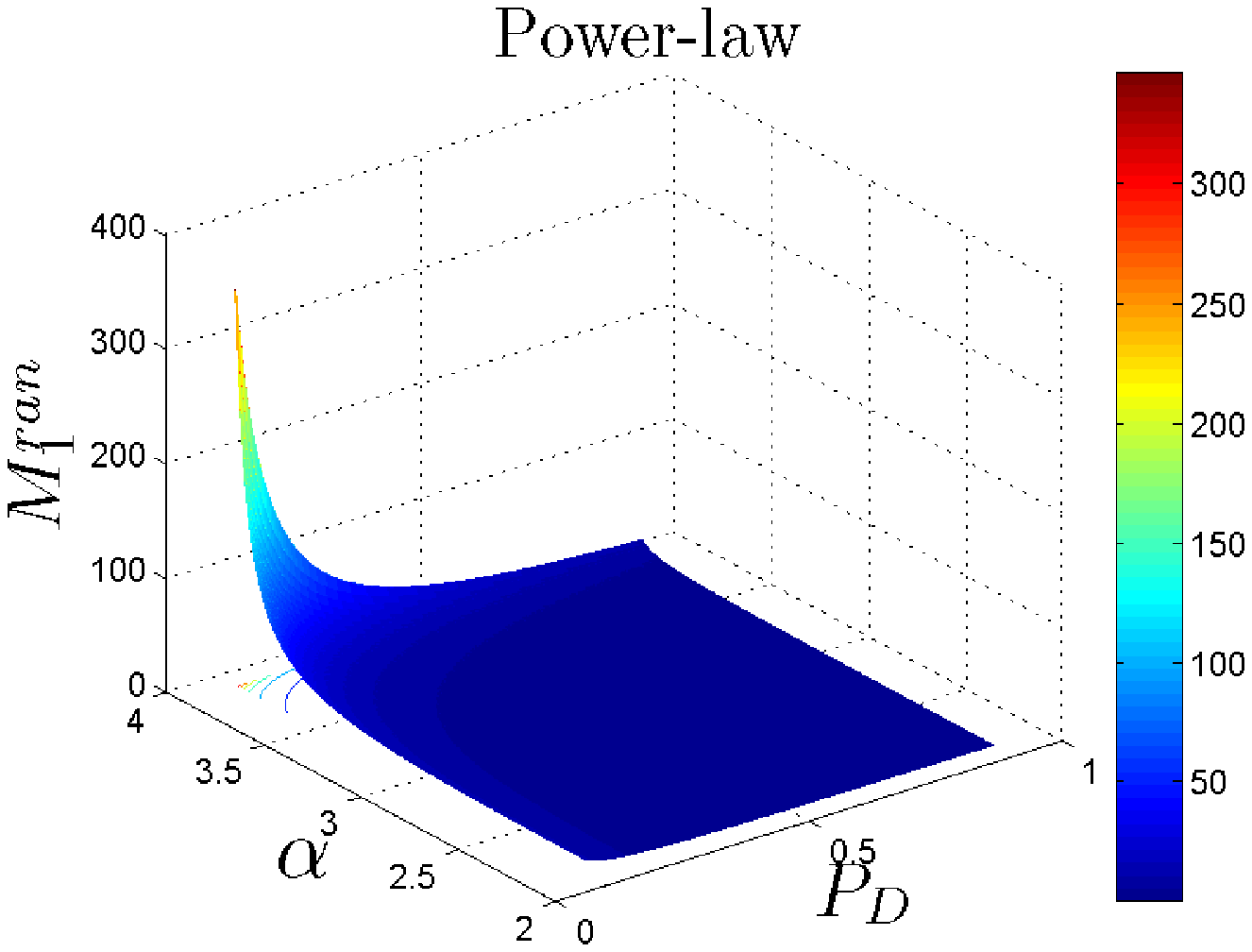}
    \caption{Performance of sequential defense against random attack in power-law networks with $P_F=0.001$.}
    \label{fig_POW_3D_6}
\end{figure}

\begin{figure}[t]
    \centering
    \includegraphics[width=3in]{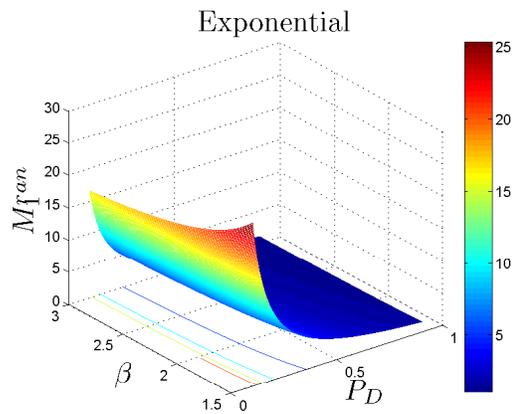}
    \caption{Performance of sequential defense against random attack in exponential networks with $P_F=0.001$.}
    \label{fig_EXP_3D_7}
\end{figure}

In this section, the proposed sequential defense mechanism is employed on canonical complex network models as well as empirical network data to evaluate the system performance and offer new insights on robust network design. The system parameters are set to be $N=10000$, $k_1=1000$, $k_N=1$, $\delta=0.01$ and $\theta=0.001$ without additional specifications.

\subsection{Critical Values of Canonical Complex Network Models}
For fair comparisons between different canonical complex network models, we set the original mean degree to be identical such that $\widehat{k}=c_1 \cdot \frac{k_1^{2-\alpha}-k_N^{2-\alpha}}{2-\alpha}=k_N+\beta$ and accordingly extract the network parameters $\widehat{k}$, $\alpha$ and $\beta$ for ER, power-law and exponential networks. As shown in Fig. \ref{fig_qc_mean_deg_2}, the critical value $q_c$ approaches to $0$ as the mean degree decreases
to $1$ for all canonical complex network models since intuitively a network is prone to disruption if every node has only one link in average. On the other hand, the critical value
increases with the mean degree as every node is able to connect to more nodes in the network in order to strengthen the network connectivity. Compared with random attack, intentional attack is shown to be more effective in disintegrating a network by sabotaging a small fraction of nodes with the highest degree. Moreover, despite the fact that the power-law network is resilient to random attack, the inherently skewed degree distribution render it quite vulnerable to intentional attack due to the existence of hub nodes with relatively high degree, which reveal the bottleneck of network robustness against intelligent attacks.

\subsection{Performance of Sequential Defense Mechanism}
 By employing the proposed sequential defense mechanism in the complex networks, we select the number of reports required to identify an attack $(M_1)$ as the performance measure for timely and efficient defense. As shown in Fig. \ref{fig_M1_ran_q_3}, $M_1^{ran}$ is shown to be a decreasing function of $P_D$ because of better precision in attack inference, and $M_1^{ran}$ increases with $P_F$ in order to distinguish attack and null attack. In addition, since $M_1^{ran}$ is also a decreasing function of $q^{ran}$, the optimal attack strategy for an intelligent adversary to disrupt the complex network would be choosing $q^{ran}=q_c^{ran}$ in order to disrupt the network while minimizing the risks of being detected. The performance of sequential defense mechanism against intentional attack is shown in Fig. \ref{fig_M1_int_PD_4}. Similar to random attack, $M_1^{int}$ increases with $P_F$ to validate the presence of attack.
Compared with random attack, the proposed sequential defense mechanism requires only a few number of reports to target intentional attack, even in the low detection probability regime.
\begin{figure}[t]
    \centering
    \includegraphics[width=3in]{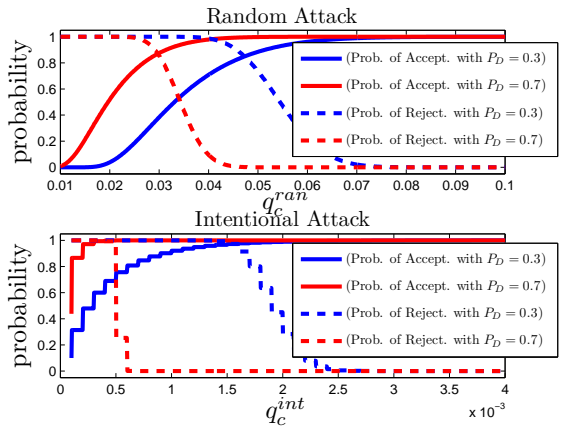}
    \caption{Probability of acceptance and rejection under worst case scenario with $P_F=0.001$. The probability of acceptance can be interpreted as the precision for attack inference, and the probability of rejection can be interpreted as the probability for an adversary to disrupt a network.}
    \label{fig_prob_q_8}
\end{figure}

\begin{figure}[t]
    \centering
    \includegraphics[width=3in]{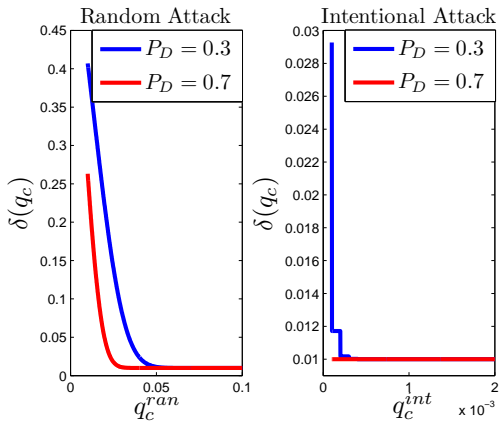}
    \caption{System level false alarm probability under worst case scenario with $P_F=0.001$ and $\delta=0.01$.}
    \label{fig_delta_q_9}
\end{figure}

\begin{figure}[t]
    \centering
    \includegraphics[width=3in]{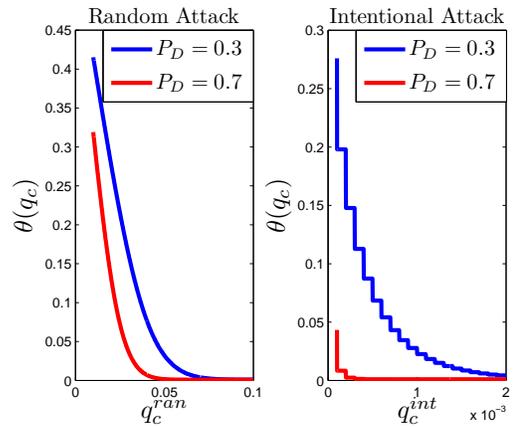}
    \caption{System level miss detection probability under worst case scenario with $P_F=0.001$ and $\theta=0.001$.}
    \label{fig_theta_q_10}
\end{figure}
To gain clear insights on the performance of the proposed sequential defense mechanism, we plot $M_1^{ran}$ and its contours with respect to the network parameters and $P_D$ for ER, power-law and exponential networks in Fig. \ref{fig_ER_3D_5}, Fig. \ref{fig_POW_3D_6} and Fig. \ref{fig_EXP_3D_7}, respectively. The network parameters are associated with the critical values under random attack as discussed in Sec. \ref{sec_attack}. For ER and exponential networks, $M_1^{ran}$ increases with the decrease of $\widehat{k}$ ($\beta$) and $P_D$ as the mean degree is proportional to $\widehat{k}$ ($\beta$) and low $P_D$ hinders the process of SPRT. For power-law networks, more skewed degree distribution (larger $\alpha$) incurs larger $M_1^{ran}$ since the network is prone to disruption as $\alpha$ increases \cite{Cohen00}.

\subsection{Reliability of Sequential Defense Mechanism}
To validate the reliability of the proposed sequential defense mechanism, the performance of worst case scenario is investigated with respect to the critical values to sustain network connectivity. In view of practical implementations, an attack decision has to be made upon the reception of $M_c=\lceil N \cdot q_c \rceil$ reports. The probability of acceptance (declaring attack) and the probability of rejection (declaring null attack) are displayed in Fig. \ref{fig_prob_q_8}. It is observed that the proposed sequential defense mechanism achieves high accuracy as the probability of acceptance (probability of rejection) approaches to $1$ ($0$) at extremely small critical values, and higher $P_D$ enhances the accuracy for attack inference, which validate that the proposed sequential defense mechanism is able to identify the attack with high precision. More importantly, given a critical value of a complex network, the probability of acceptance can be interpreted as the precision of identifying an attack prior to the network disruption, and the probability of rejection can be interpreted as the probability for an adversary to disrupt a complex network.
The system level parameters ($\delta(q_c),\theta(q_c)$) of the worst case scenario are demonstrated in Fig. \ref{fig_delta_q_9} and Fig. \ref{fig_theta_q_10}, respectively. These parameters converge to the desired system level parameters ($\delta,\theta$) at extremely small critical values, suggesting that the proposed sequential defense mechanism offers reliable and effective approaches against random and intentional attacks in complex networks.
\begin{figure}[t]
    \centering
    \includegraphics[width=3.5in]{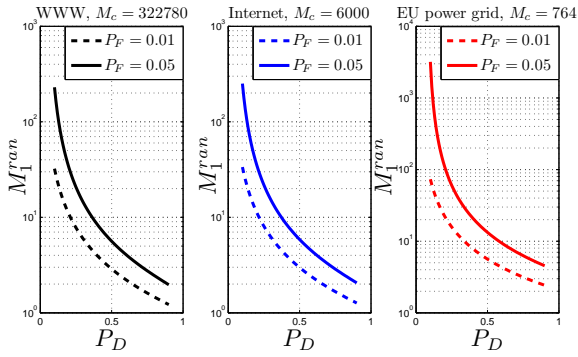}
    \caption{Performance of sequential defense mechanism under random attack with empirical network data. The critical values are $(q_c^{ran},M_c)=(0.9909,322780)$, $(q_c^{ran},M_c)=(0.9673,6000)$ and $(q_c^{ran},M_c)=(0.629,764)$ for the WWW, Internet and EU power grid, respectively.}
    \label{fig_empirical_random_11}
\end{figure}

\begin{figure}[t]
    \centering
    \includegraphics[width=3in]{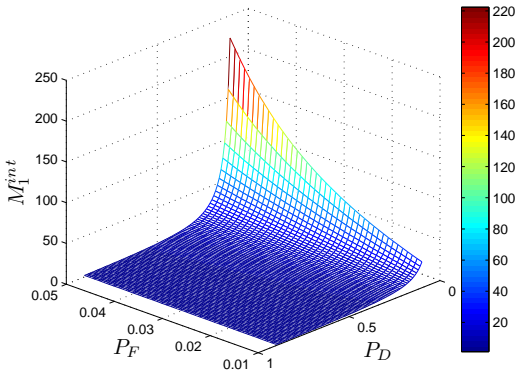}
    \caption{Performance of sequential defense mechanism under intentional attack with empirical network data. The critical values are $(q_c^{int},M_c)=(0.067,21824)$, $(q_c^{int},M_c)=(0.03,187)$ and $(q_c^{int},M_c)=(0.275,766)$ for the WWW, Internet and EU power grid, respectively.}
    \label{fig_empirical_intentional_12}
\end{figure}

\subsection{Empirical Network Data}
As a demonstration, the proposed sequential defense mechanism are implemented in real-world large-scale networks with network parameters extracted from empirical network data collected in \cite{Albert00,Sole08}. The WWW contains $325729$ nodes (webpages) and $1798353$ links with $\mathbb{E}[\mathbf{K}_0]=4.6$. The Internet router-level map contains $6209$ nodes (routers) and $12200$ links with $\mathbb{E}[\mathbf{K}_0]=3.4$. The EU power grid contains $2783$ nodes (power stations) and $3762$ links with $\mathbb{E}[\mathbf{K}_0]=3.4$. The WWW and the Internet are power-law networks with network parameters $\alpha=2.1$ and $\alpha=2.5$, respectively. The EU power grid is an exponential network with network parameter $\beta=1.63$. As shown in Fig. \ref{fig_empirical_random_11}, the number of reports required to identify random attack ($M_1^{ran}$) is lower than the threshold $M_c$, even in the case of weak detection capability (low $P_D$). On the other hand, $M_1^{ran}$ increases with $P_F$ as the data fusion center requires more reports to distinguish between attack and null attack
when the false alarm probability increases. Fig. \ref{fig_empirical_intentional_12} displays the performance of the sequential defense mechanism with respect to $P_D$ and $P_F$ under intentional attack. A surge increase of $M_1^{int}$ is observed with the decrease of $P_D$ and the increase of $P_F$, suggesting that the defense configurations have to be adjusted according to network characteristics in order to guarantee robust and reliable operations of the entire system, especially for the networks which are particularly vulnerable to intentional attack.

\subsection{US Power Grid}
\begin{figure}[t]
    \centering
    \includegraphics[width=3.5in]{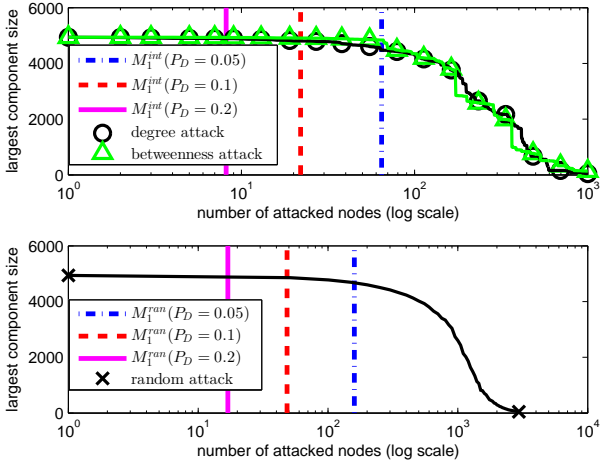}
    \caption{Performance of sequential defense mechanism on US power grid topology \cite{Watts98} under different attack schemes with $P_F=0.005$. The power grid topology contains $4941$ nodes (power stations) and $6594$ edges (power lines). For random attack, the results are averaged over $100$ realizations. The expected number of reports ($M_1$) needed for attack detection is relatively small and it decreases as $P_D$ increases. The proposed sequential defense mechanism is quite effective in the sense the network suffers slight connectivity loss when the number of attacked nodes is less than $M_1$ (i.e., the undetectable region), even for small $P_D$.}
    \label{fig_Power_Grid_13}
\end{figure}

We implement the proposed sequential defense mechanism on the US power grid topology collected in \cite{Watts98}. In addition to degree and random attacks, we also consider betweenness attack, where betweenness of a node is defined as the fraction of all shortest paths passing through the node among all shortest paths between each node pair in the network \cite{Freeman77}. As shown in Fig. \ref{fig_Power_Grid_13}, the network resilience is evaluated in terms of the largest component size when a subset of nodes is removed from the network. Given the expected number of reports required for attack detection $M_1$, if an adversary attacks less than $M_1$ nodes in the network, then the attack will not be detected, which we refer to as the undetectable region. As shown in Fig. \ref{fig_Power_Grid_13}, $M_1$ decreases as $P_D$ increases, and it is shown to be relatively small compared with the network size. Notably, in the undetectable region, most of the nodes are still connected, even with small $P_D$. The results indicate that the proposed sequential defense mechanism is quite effective in attack detection and the network suffers slight connectivity loss in the undetectable region. Note that the perfect protection defense strategy proposed in \cite{Xiao11} is a degenerate case of our proposed mechanism when $P_D \rightarrow 1$ and $P_F \rightarrow 0$. It results in extremely small $M_1$ and suggests that a network can be robust to attacks if perfect protection is plausible.

\begin{figure}[t]
    \centering
    \includegraphics[width=3in]{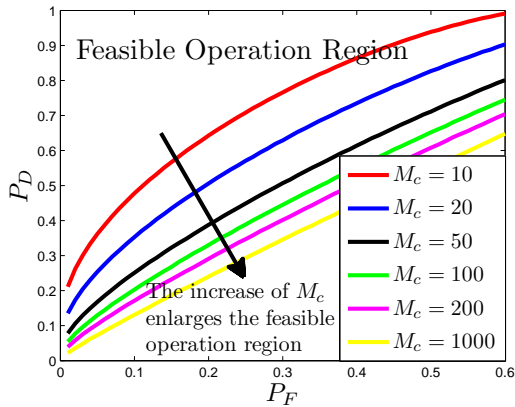}
    \caption{Operation curves of $P_D$ and $P_F$ with respect to a specified network disruption threshold $M_c$ under intentional attack. The feasible operation region are composed of the feasible parameters ($P_D$,$P_F$) such that $M_c \geq M_1^{int}$ in order to guarantee network robustness.}
    \label{fig_robust_network_14}
\end{figure}
\section{Discussions toward Robust Network Design}
\label{sec_robust}
Based on the performance analysis of a complex network empowered with the proposed sequential defense mechanism, we provide some insights on robust network design against attacks in complex networks. To guarantee that the process of SPRT terminates before an adversary paralyzes the entire system, the baseline requirement for the network disruption threshold is $M_c \geq \max \left\{ M_1^{ran},M_1^{int} \right\}$, which ensures that the data fusion center can acquire sufficient information for attack inference and make immediate reactions against the attacks prior to the network disruption. In other words, in the network operator's point of view, one has to enhance the critical values of a network under attacks to maintain the operations of the defense mechanism in the system, which can be achieved via different approaches in consideration of the network configurations and implementation costs. Consequently, this paper offers analytically tractable tools for robust network design and network defense performance assessment. Potential approaches to network robustness enhancement are discussed as follows.

\begin{itemize}
  \item \textbf{Link Addition}.
  As shown in Fig. \ref{fig_qc_mean_deg_2}, adding more links in the network (i.e., increasing the mean degree) strengthens the network connectivity and thereby offers more protection against attacks. Although link addition is a straightforward solution \cite{Ghosh06,Yehezkel12}, the major drawback of link addition is that it may decrease the system revenue if the costs for link constructions are high, such as the transportation systems.
  \item \textbf{Topology Adjustment}.
  As the network resilience varies from network parameters, the critical value of a network can be modified by topology adjustment while keeping the number of links in the network (i.e., the mean degree) unchanged \cite{Moreira09}, which is especially suitable for networks connected by logical configurations (e.g., the WWW). For an example, as demonstrated in Fig. \ref{fig_qc_mean_deg_2}, adjusting a power-law network to an exponential network enhances the resilience against intentional attack at the cost of decreasing the resilience against random attack, which offers tradeoffs between $M_1^{ran}$ and $M_1^{int}$.
  \item \textbf{Detection Capability Enhancement}.
  In cases that link addition and topology adjustment are infeasible and thereby the critical values can not be modified, one has to enhance the detection capability to provide reliable network defense against attacks. Take sequential defense against intentional attack as a motivating example, the sequential defense mechanism is able to target intentional attack if the network disruption threshold $M_c$ is no less than $M_1^{int}$. Applying this criterion to (\ref{eqn_performance_intentional_M1}), the feasible parameters $P_D$ and $P_F$ for sequential defense mechanism need to satisfy the inequality
\begin{align}
\label{eqn_robust_network}
&&P_D\log \frac{P_D}{P_F} + (1-P_D)\log \frac{1-P_D}{1-P_F} \nonumber \\
&& \geq \frac{\theta \log \frac{\theta}{1-\delta}+(1-\theta) \log \frac{1-\theta}{\delta}}{M_c}.
\end{align}
The operation curves when the equality in (\ref{eqn_robust_network}) holds given a specified network disruption threshold ($M_c$) are shown in Fig. \ref{fig_robust_network_14}, which can be interpreted as the minimum detection probability ($P_D$) required to perform sequential defense with respect to a false alarm probability $P_F$ and $M_c$. The feasible operation region is composed of the parameters ($P_D$,$P_F$) satisfying the inequality in (\ref{eqn_robust_network}), and the increase of $M_c$ enlarges the feasible operation region since the data fusion center can acquire more reports for attack inference prior to network disruption, even in the low $P_D$ regime.
\end{itemize}

\section{conclusion}
\label{sec_con}
In this paper, a sequential defense mechanism based on sequential hypothesis test is proposed in complex networks with an aim of enhancing the network robustness of networked engineering systems. This mechanism provides timely and efficient defense against random and intentional attacks by sequentially acquiring binary attack status of each node in descending degree order. The data collection process terminates once a final decision has been made by the data fusion center, which is particularly preferable in networking paradigms with stringent data transmission resources. Therefore the low computation complexity and sequential transmission schemes render this defense mechanism compatible to practical networked engineering systems. A parametric plug-in model is proposed to evaluate the performance of the proposed sequential defense mechanism.
By implementing this mechanism on the canonical complex network models as well as the empirical network data extracted from the WWW, the Internet, the EU power grid, and the US power grid topology, the results validate the effectiveness and reliability of this mechanism against fatal attacks. These attacks can be identified with high precision with limited binary attack status reported from a small subset of nodes in the network
and thereby immediate defense reactions can be performed prior to the network disruption, even in the weak topological vulnerability and low detection capability regime. Based on the performance analysis and network configurations, several approaches including link addition, topology adjustment and detection capability enhancement are elucidated to guarantee robust operations of the entire system.
Consequently, this paper provides profound theoretic framework of sequential defense in complex networks and offers new insights on robust network design in complex networks.

\bibliographystyle{apsrev4-1}
\bibliography{IEEEabrv,sequential}

\clearpage 

\appendix
\section{Critical Value for Random Attack}
\label{appex_critical_random}
Following \cite{Cohen00}, given the original degree distribution $P_0(k_0)$, the new degree distribution of the network after randomly removing $q$ fraction of nodes (the links emanating from the nodes are removed as well) is
\begin{align}
\label{eqn_random_removal}
P(k)=\sum_{k_0=k}^{k_{max}} P_0(k_0)\binom{k_0}{k}(1-q)^k q^{k_0-k}.
\end{align}
Applying (\ref{eqn_random_removal}) to (\ref{eqn_condition}), the criterion for the percolation-based connectivity after random attack becomes
\begin{align}
\label{eqn_random_removal_critical}
\frac{(1-q)^2 \mathbb{E}[\mathbf{K_0}^2]+q(1-q)\mathbb{E}[\mathbf{K_0}]}{(1-q) \mathbb{E}[\mathbf{K_0}]}=2.
\end{align}
Reorganizing (\ref{eqn_random_removal_critical}), we obtain the critical value $q_c^{ran}=1-\frac{1}{\tau_0-1}$ as in (\ref{eqn_random_attack_thre}).
For ER network, we have $\mathbb{E}[\mathbf{K_0}]=\widehat{k}$ and $\mathbb{E}[\mathbf{K_0}^2]=\widehat{k}^2+\widehat{k}$, applying to (\ref{eqn_random_attack_thre}), we have
\begin{align}
\label{eqn_random_attack_ER}
q_{c}^{ran-ER}=1-\frac{1}{\widehat{k}}.
\end{align}
For power-law network, the $r$th moment of the degree distribution is $\mathbb{E}[\mathbf{K}^r]=c_1 \cdot \frac{k_{max}^{r-\alpha+1}- k_{min}^{r-\alpha+1}}{r-\alpha+1}$, where $c_1=\frac{1-\alpha}{k_{max}^{1-\alpha}-k_{min}^{1-\alpha}}$ is the normalization factor. Applying to (\ref{eqn_random_attack_thre}), we obtain
\begin{align}
\label{eqn_random_attack_Power}
q_{c}^{ran-POW}=1-\frac{1}{\frac{2-\alpha}{3-\alpha}\frac{k_{1}^{3-\alpha}- k_{N}^{3-\alpha}}{k_{1}^{2-\alpha}- k_{N}^{2-\alpha}}-1}.
\end{align}
For exponential network, $P(k)=c_2 \cdot \frac{1}{\beta}e^{-\frac{k}{\beta}}$, where $c_2=\frac{1}{e^{-\frac{k_N}{\beta}}-e^{-\frac{k_1}{\beta}}}$ is the normalization factor. In the large scale network limit (i.e., $k_1 \rightarrow \infty$), $c_2=e^{\frac{k_N}{\beta}}$, $\mathbb{E}[\mathbf{K}]=k_N+\beta$ and $\mathbb{E}[\mathbf{K}^2]=k_N^2+2 k_N \beta+2 \beta^2$. We obtain
\begin{align}
\label{eqn_random_attack_Exp}
q_c^{ran-EXP}=1-\frac{1}{\frac{k_N^2+2 k_N \beta+2 \beta^2}{k_N+\beta}-1}.
\end{align}

\section{Critical Value for Intentional Attack}
\label{appex_critical_intentional}
For simplicity, we derive the critical value under intentional attack in the large scale network limit ($k_1 \rightarrow \infty$) as follows. It is also suggested in \cite{Cohen01} the large scale network limit assumption has negligible impacts on the accuracy of the critical value provided that the network size ($N$) is large enough. The methodology for deriving the critical value under intentional attack is to specify the relations between the cutoff degree $\widetilde{d}_{max}$ and the fraction of removed nodes $q$ using (\ref{eqn_intentional_removal}), and then apply the cutoff degree to the deletion probability $\widetilde{q}$ and the criterion for percolation-based connectivity in (\ref{eqn_random_attack_thre}) to obtain the critical value $q_c^{int}$.

For ER network, from (\ref{eqn_intentional_removal}) we have $e^{-\widehat{k}} \frac{\widehat{k}^{k_1}}{k_1!}=\frac{1}{N}$ and $q=\sum_{k=\widetilde{k}_{max}}^{k_1} e^{-\widehat{k}} \frac{\widehat{k}^k}{k!}=1-\frac{\Gamma(\widetilde{k}_{max},\widehat{k})}{\left( \widetilde{k}_{max}-1 \right)!}$, where $\Gamma(s,x)=\int_x^{\infty} t^{s-1} e^{-t} dt$ is the upper incomplete gamma function.
The deletion probability becomes
$\widetilde{q}=\sum_{k=\widetilde{k}_{max}}^{k_1} \frac{k P(k)}{\mathbb{E}[\mathbf{K_0}]} =\sum_{k=\widetilde{k}_{max}}^{k_1} e^{-\widehat{k}} \frac{\widehat{k}^{k-1}}{(k-1)!}= q-\frac{1}{N}+ e^{-\widehat{k}} \frac{\widehat{k}^{\widetilde{k}_{max}-1}}{\left( \widetilde{k}_{max}-1 \right)!}$, and the cutoff degree $\widetilde{k}_{max}$ can be obtained by solving $\widetilde{q}=1-\frac{1}{\widehat{k}}$. Consequently, the critical value under intentional attack is
\begin{align}
\label{eqn_intentional_attack_ER}
q_c^{int-ER}=\frac{1}{N}- e^{-\widehat{k}} \frac{\widehat{k}^{\widetilde{k}_{max}-1}}{\left( \widetilde{k}_{max}-1 \right)!}+1-\frac{1}{\widehat{k}}.
\end{align}

For power-law network, from (\ref{eqn_intentional_removal}) we have $\widetilde{k}_{max}=k_N \left( q+\frac{1}{N} \right)^{\frac{1}{1-\alpha}}$ and $q\overset{N \rightarrow \infty}{=}\left( \frac{\widetilde{k}_{max}}{k_N}\right)^{1-\alpha}$ for all $\alpha > 1$. The relations between $\widetilde{q}$ and $q$ are $\widetilde{q}=\int_{k=\widetilde{k}_{max}}^{k_1} \frac{k P(k)}{\mathbb{E}[\mathbf{K_0}]} d k= \left(\frac{\widetilde{k}_{max}}{k_N}\right)^{2-\alpha}=q^{\frac{2-\alpha}{1-\alpha}}$. The cutoff degree $\widetilde{k}_{man}$ can be solved by applying the link deletion probability to the criterion in (\ref{eqn_random_attack_thre}), which yields the equation $\left(\frac{\widetilde{k}_{max}}{k_N}\right)^{2-\alpha}-k_N \left(\frac{2-\alpha}{3-\alpha}\right) \left[ \left(\frac{\widetilde{k}_{max}}{k_N}\right)^{3-\alpha}-1 \right]-2=0$ \cite{Cohen01}, and we thereby obtain
\begin{align}
\label{eqn_intentional_attack_Power}
q_c^{int-POW}=\left(\frac{\widetilde{k}_{max}}{k_N}\right)^{1-\alpha}.
\end{align}

For exponential network, the relations between the cutoff degree and the fraction of removed nodes are $\widetilde{k}_{max}=-\beta \ln \left( q+\frac{1}{N}\right) + k_N$. The deletion probability becomes $\widetilde{q}=\int_{k=\widetilde{k}_{max}}^{k_1} \frac{k P(k)}{\mathbb{E}[\mathbf{K_0}]} d k=\frac{\exp\left(\frac{k_N}{\beta}\right)}{k_N+\beta} \cdot \left( \widetilde{k}_{max}+\beta \right)\exp \left(-\frac{\widetilde{k}_{max}} {\beta}\right)$. If $k_N$ is negligible (i.e., $k_N=0$), we have $\widetilde{q}=\left[ 1-\ln \left( q+ \frac{1}{N}\right) \right]\left(q+ \frac{1}{N} \right)$. Applying the result to (\ref{eqn_random_attack_Exp}), the critical value under intentional attack can be obtained by solving
\begin{align}
\label{eqn_intentional_attack_Exp}
&\left[ 1-\ln \left( q_c^{int-EXP}+ \frac{1}{N}\right) \right]\left(q_c^{int-EXP}+ \frac{1}{N}\right) \nonumber\\
&+\frac{k_N+\beta}{k_N^2+2 k_N \beta+2 \beta^2-k_N-\beta}-1=0.
\end{align}
\clearpage

\end{document}